\documentclass[pre,twocolumn,aps]{revtex4}
\newcommand{\bec}[1]{\mbox{\boldmath $ #1$}}
\usepackage{graphicx}
\begin{document}
\leftline{\bf Nonlinear Processes in Geophysics, v. 11, 343-350
(2004)}
\bigskip
\title{TURBULENT THERMAL DIFFUSION OF AEROSOLS
IN GEOPHYSICS AND IN LABORATORY EXPERIMENTS}
\author{Alexander Eidelman}
\author{Tov Elperin}
\author{Nathan Kleeorin}
\author{Alexander Krein}
\author{Igor Rogachevskii}
\affiliation{The Pearlstone Center for Aeronautical Engineering
Studies, Department of Mechanical Engineering, The Ben-Gurion
University of the Negev, POB 653, Beer-Sheva 84105, Israel}
\author{Julia Buchholz}
\author{Gerd Gr\"{u}nefeld}
\affiliation{Faculty of Mechanical Engineering, RWTH Aachen
University, Aachen, Germany}
\date{\today}

\begin{abstract}
We discuss a new phenomenon of turbulent thermal diffusion
associated with turbulent transport of aerosols in the atmosphere
and in laboratory experiments. The essence of this phenomenon is
the appearance of a nondiffusive mean flux of particles in the
direction of the mean heat flux, which results in the formation of
large-scale inhomogeneities in the spatial distribution of
aerosols that accumulate in regions of minimum mean temperature of
the surrounding fluid. This effect of turbulent thermal diffusion
was detected experimentally. In experiments turbulence was
generated by two oscillating grids in two directions of the
imposed vertical mean temperature gradient. We used Particle Image
Velocimetry to determine the turbulent velocity field, and an
Image Processing Technique based on an analysis of the intensity
of Mie scattering to determine the spatial distribution of
aerosols. Analysis of the intensity of laser light Mie scattering
by aerosols showed that aerosols accumulate in the vicinity of the
minimum mean temperature due to the effect of turbulent thermal
diffusion.
\end{abstract}

\maketitle

\section{Introduction}
\label{sec:intro}

Aerosols are a universal feature of the Earth's atmosphere. They
can significantly affect the heat balance and dynamics of the
atmosphere, climate, atmospheric chemistry, radiative transport
and precipitation formation (see, e.g., Twomey, 1977; Seinfeld,
1986; Flagan and Seinfeld, 1988; Pruppacher and Klett 1997;
Lohmann and Lesins, 2002; Kaufman et al. 2002; Anderson et al.
2003; and references therein). Formation of aerosol clouds is of
fundamental significance in many areas of environmental sciences,
physics of the atmosphere and meteorology (see, e.g., Seinfeld,
1986; Flagan and Seinfeld, 1988; Paluch and Baumgardner, 1989;
Shaw et al. 1998; Shaw 2003; and references therein). It is well
known that turbulence causes decay of inhomogeneities in spatial
distribution of aerosols due to turbulent diffusion (see, e.g.,
Csanady, 1980; McComb 1990; Stock 1996), whereas the opposite
effect, the preferential concentration of aerosols in atmospheric
turbulence is not well understood.

Elperin et al. (1996; 1997; 1998; 2000b; 2001) recently found that
in a turbulent fluid flow with a nonzero mean temperature gradient
an additional mean flux of aerosols appears in the direction
opposite to that of the mean temperature gradient, which is known
as the phenomenon of turbulent thermal diffusion. This effect
results in the formation of large-scale inhomogeneities in the
spatial distribution of the aerosol particles.

The phenomenon of turbulent thermal diffusion is important for
understanding atmospheric phenomena (e.g., atmospheric aerosols,
smog formation, etc.). There exists a correlation between the
appearance of temperature inversions and the aerosol layers
(pollutants) in the vicinity of the temperature inversions (see,
e.g., Seinfeld, 1986; Flagan and Seinfeld, 1988). Moreover,
turbulent thermal diffusion can cause the formation of large-scale
aerosol layers in the vicinity of temperature inversions in
atmospheric turbulence (Elperin et al. 2000a; 2000c).

The main goal of this paper is to describe the experimental
detection of a new phenomenon of turbulent thermal diffusion. To
accomplish this, we constructed an oscillating grids turbulence
generator (for details see Eidelman et al. 2002). Recent studies
by De Silva and Fernando (1994); Srdic et al. (1996), Shy et al.
(1997) have demonstrated the feasibility of generating nearly
isotropic turbulence by two oscillating grids. Turbulent diffusion
in oscillating grids turbulence was investigated by Ott and Mann
(2000). In order to study turbulent thermal diffusion we used
Particle Image Velocimetry to characterize a turbulent velocity
field, and an Image Processing Technique based on Mie scattering
to determine the spatial distribution of the particles. Our
experiments were performed in two directions of the vertical mean
temperature gradient: an upward mean temperature gradient (formed
by a cold bottom and a hot top wall of the chamber) and a downward
mean temperature gradient. We found that in a flow with a downward
mean temperature gradient, particles accumulate in the vicinity of
the top wall of the chamber (whereby the mean fluid temperature is
minimal), and in a flow with an upward mean temperature gradient
particles accumulate in the vicinity of the bottom wall of the
chamber due to the effect of turbulent thermal diffusion.

The paper is organized as follows. Section II discusses the
physics of turbulent thermal diffusion, and Section III describes
the experimental set-up for a laboratory study of this effect. The
experimental results are presented in Section IV, and a detailed
analysis of experimental detection of turbulent thermal diffusion
is performed in Section V. Finally, conclusions are drawn in
Section VI.

\section{Turbulent thermal diffusion}
\label{sec:ttd}

Evolution of the number density $ n(t,{\bf r}) $ of small
particles in a turbulent flow is determined by
\begin{eqnarray}
{\partial n \over \partial t} + \bec{\nabla} \cdot(n {\bf v}_p) =
- \bec{\nabla} \cdot {\bf J}_{_{M}} \;, \label{A1}
\end{eqnarray}
where $ {\bf J}_{_{M}} $ is the molecular flux of the particles
and ${\bf v}_p$ is the velocity of the particles in the turbulent
fluid velocity field. Averaging Eq.~(\ref{A1}) over the statistics
of the turbulent velocity field yields the following equation for
the mean number density of particles $ \bar N \equiv \langle n
\rangle :$
\begin{eqnarray}
&& {\partial \bar N \over \partial t} + \bec{\nabla} \cdot (\bar N
\bar{\bf V}_p) = - \bec{\nabla} \cdot (\bar{\bf J}_{_{T}} +
\bar{\bf J}_{_{M}}) \;,
\label{A2}\\
&& \bar{\bf J}_{_{T}} = \bar N {\bf V}_{\rm eff} - D_{_{T}}
\bec{\nabla} \bar N \;,
\label{A3}
\end{eqnarray}
where $ D_{_{T}} = (\tau /3) \langle {\bf u}^2 \rangle $ is the
turbulent diffusion coefficient,  $\tau$ is the momentum
relaxation time of the turbulent velocity field, $ {\bf v}_p =
\bar{\bf V}_p + {\bf u} ,$ $\, \bar{\bf V}_p = \langle {\bf v}_p
\rangle $ is the mean particle velocity, and the effective
velocity is
\begin{eqnarray}
{\bf V}_{\rm eff} = - \langle \tau {\bf u} (\bec{\nabla} \cdot
{\bf u}) \rangle \; . \label{P3}
\end{eqnarray}
Here $ \bar{\bf J}_{_{M}} = - D (\bec{\nabla} \bar N + k_{t}
\bec{\nabla} \bar T / \bar T) $ is the mean molecular flux of
particles, $ D $ is the coefficient of molecular diffusion, $k_t$
is the thermal diffusion ratio, and $\bar T = \langle T \rangle $
is the mean fluid temperature. Equations~(\ref{A2}) and~(\ref{A3})
were previously derived by different methods (see Elperin et al.
1996; 1997; 1998; 2000b; 2001; Pandya and Mashayek 2002).

For noninertial particles advected by a turbulent fluid flow,
particle velocity $ {\bf v}_p $ coincides with fluid velocity $
{\bf v} ,$ and $\bec{\nabla} \cdot {\bf v} \approx - ({\bf v}
\cdot \bec{\nabla}) \rho / \rho \approx ({\bf v} \cdot
\bec{\nabla}) T / T ,$ where $ \rho $ and $T$ are the density and
temperature of the fluid. Thus, the effective velocity~(\ref{P3})
for noninertial particles can be given by $ {\bf V}_{\rm eff} = -
D_{_{T}} (\bec{\nabla} \bar T) / \bar T ,$ which takes into
account the equation of state for an ideal gas but neglects small
gradients of the mean fluid pressure.

For inertial particles, velocity $ {\bf v}_p $ depends on the
velocity of the surrounding fluid $ {\bf v} $. Velocity $ {\bf
v}_p $ can be determined by the equation of motion for a particle.
Solving the equation of motion for a small solid particle with $
\rho_p \gg \rho $ yields: ${\bf v}_p = {\bf v} - \tau_p d{\bf v} /
dt + {\rm O}(\tau_p^2)$ (see Maxey 1987), where $ \tau_p $ is the
Stokes time and $ \rho_p $ is the material density of the
particles. In that case
\begin{eqnarray}
\bec{\nabla} \cdot {\bf v}_p = \bec{\nabla} \cdot {\bf v} + \tau_p
{\Delta P \over \rho} + {\rm O}(\tau_p^2) \;, \label{P6}
\end{eqnarray}
and the effective velocity of the inertial particles can be given
by $ {\bf V}_{\rm eff} = - D_{_{T}} (1 + \kappa) (\bec{\nabla}
\bar T) / \bar T ,$ where $P$ is the fluid pressure. Coefficient $
\kappa $ depends on particle inertia $(m_p / m_\mu),$ the
parameters of turbulence (Reynolds number) and the mean fluid
temperature (Elperin et al. 1996; 1997; 1998; 2000b; 2001). Here $
m_p  $ is the particle mass and $ m_\mu $ is the mass of molecules
of the surrounding fluid. The turbulent flux of
particles~(\ref{A3}) can be rewritten as
\begin{eqnarray}
\bar{\bf J}_{_{T}} = - D_{_{T}} \biggl[k_{_{T}} {(\bec{\nabla}
\bar T) \over \bar T} + \bec{\nabla} \bar N \biggr] \;, \label{P7}
\end{eqnarray}
where $k_{_{T}} = (1 + \kappa) \bar N $. The first term in the RHS
of Eq.~(\ref{P7}) describes turbulent thermal diffusion, while the
second term in the turbulent flux of particles~(\ref{P7})
describes turbulent diffusion. Parameter $k_{_{T}}$ can be
interpreted as the turbulent thermal diffusion ratio, and $
D_{_{T}} k_{_{T}} $ is the coefficient of turbulent thermal
diffusion. Turbulent thermal diffusion causes the formation of a
large-scale pattern wherein the initial spatial distribution of
particles in a turbulent fluid flow evolves into a large-scale
inhomogeneous distribution, i.e., particles accumulate in the
vicinity of the minimum mean temperature of the surrounding fluid
(Elperin et al. 1996; 1997; 1998; 2000b; 2001).

The mechanism responsible for the occurrence of turbulent thermal
diffusion for particles with $\rho_p \gg \rho$ can be described as
follows. Inertia causes particles inside the turbulent eddies to
drift out to the boundary regions between the eddies (i.e.,
regions with low vorticity or high strain rate and maximum of
fluid pressure). Thus, particles accumulate in regions with
maximum pressure of the turbulent fluid. For simplicity, let us
consider a pure inertial effect, i.e., we assume that
$\bec{\nabla} \cdot {\bf v} = 0$. This inertial effect results in
$ \bec{\nabla} \cdot {\bf v}_p \propto \tau_p \Delta P \not= 0 .$
On the other hand, Eq.~(\ref{A1}) for large Peclet numbers yields
$ \bec{\nabla} \cdot {\bf v}_p \propto - dn / dt .$ The latter
formula implies that $dn / dt \propto - \tau_p \Delta P $, i.e.,
inertial particles accumulate ($ dn / dt > 0)$ in regions with
maximum pressure of the turbulent fluid (where $ \Delta P < 0 ) $.
Similarly, there is an outflow of particles from regions with
minimum pressure of fluid. In homogeneous and isotropic turbulence
without large-scale external gradients of temperature, a drift
from regions with increased or decreased concentration of
particles by a turbulent flow of fluid is equiprobable in all
directions, and pressure and temperature of the surrounding fluid
do not correlate with the turbulent velocity field. Thus only
turbulent diffusion of particles takes place.

In a turbulent fluid flow with a mean temperature gradient, the
mean heat flux $\langle {\bf u} \Theta \rangle$ is not zero, i.e.,
the fluctuations of fluid temperature $\Theta=T-\bar T$ and the
velocity of the fluid correlate. Fluctuations of temperature cause
fluctuations of fluid pressure. These fluctuations result in
fluctuations of the number density of particles. Indeed, an
increase in pressure of the surrounding fluid is accompanied by an
accumulation of particles. Therefore, the direction of the mean
flux of particles coincides with that of the heat flux, $\langle
{\bf v}_p n \rangle \propto \langle {\bf u} \Theta \rangle \propto
- \bec{\nabla} \bar T ,$ i.e., the mean flux of particles is
directed to the area with minimum mean temperature, and the
particles accumulate in that region.

\section{Experimental set-up}
\label{sec:setup}

In this Section we investigate experimentally the effect of
turbulent thermal diffusion. The experiments were conducted in an
oscillating grids turbulence generator in air flow (see Fig.~1).
The test section consisted of a rectangular chamber of dimensions
$29 \times 29 \times 58$ cm (see Fig.~2). Pairs of vertically
oriented grids with bars arranged in a square array were attached
to the right and left horizontal rods. Both grids were driven
independently with speed-controlled motors. The grids were
positioned at a distance of two-grid meshes from the chamber walls
parallel to them. A two-grid system can oscillate at a
controllable frequency up to $20$ Hz. The grid stroke was adjusted
within a range of $1$ to $10$ cm.

\begin{figure}
\vspace*{2mm}
\centering
\includegraphics[width=8cm]{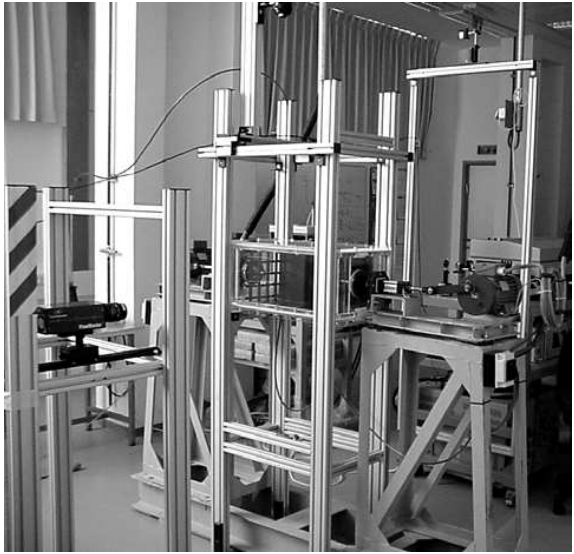}
\caption{\label{Fig1} The experimental set-up.}
\end{figure}

\begin{figure}
\vspace*{2mm}
\centering
\includegraphics[width=8cm]{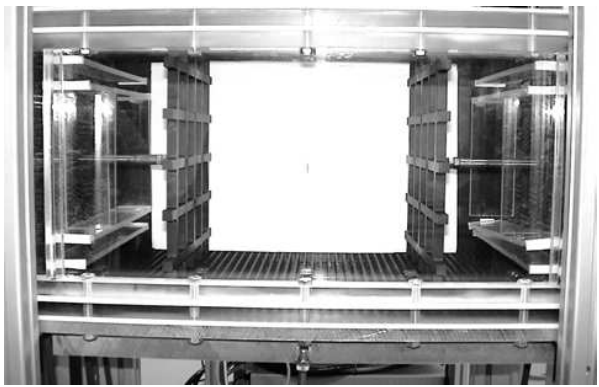}
\caption{\label{Fig2} The test section of the oscillating grids
turbulence generator.}
\end{figure}

A vertical mean temperature gradient in the turbulent flow was
formed by attaching two aluminium heat exchangers to the bottom
and top walls of the test section (see Fig.~2). The experiments
were performed in two directions of the mean temperature gradient:
an upward mean temperature gradient (a cold bottom and a hot top
wall of the chamber) and a downward mean temperature gradient (a
heated bottom and a cold top wall of the chamber). In order to
improve the heat transfer in the boundary layers at the walls we
used a heat exchanger with rectangular pins $3 \times 3 \times 15$
mm. This allowed us to support a mean temperature gradient in the
core of the flow up to 200 K/m in the upward direction and up to
110 K/m in the downward direction at a mean temperature of about
300 K. The temperature was measured with a high-frequency response
thermocouple.

The velocity field was measured using a Particle Image Velocimetry
(PIV), see Raffel  et al. (1998). A digital PIV system with
LaVision Flow Master III was used. A double-pulsed light sheet was
provided by a Nd-YAG laser source (Continuum Surelite $ 2 \times
170$ mJ). The light sheet optics includes spherical and
cylindrical Galilei telescopes with tuneable divergence and
adjustable focus length. We used a progressive-scan 12 bit digital
CCD camera (pixel size $6.7 \, \mu$m $\times 6.7 \, \mu$m each)
with a dual-frame-technique for cross-correlation processing of
captured images. A programmable Timing Unit (PC interface card)
generated sequences of pulses to control the laser, camera and
data acquisition rate. The software package DaVis 6 was applied to
control all hardware components and for 32 bit image acquisition
and visualization. This software package contains PIV software for
calculating the flow fields using cross-correlation analysis.
Velocity maps and their characteristics, e.g., statistics and PDF,
were analyzed with this package plus an additional developed
software package. An incense smoke with sub-micron particles (with
$ \rho_p / \rho \sim 10^3) $ as a tracer was used for the PIV
measurements. Smoke was produced by high temperature sublimation
of solid incense particles. Analysis of smoke particles using a
microscope (Nikon, Epiphot with an amplification of 560) and a
PM-300 portable laser particulate analyzer showed that these
particles have a spherical shape and that their mean diameter is
$0.7 \mu$m.

Mean and r.m.s. velocities, two-point correlation functions and an
integral scale of turbulence from the measured velocity fields
were determined. A series of 100 pairs of images acquired with a
frequency of 4 Hz were stored for calculating the velocity maps
and for ensemble and spatial averaging of turbulence
characteristics. The center of the measurement region coincides
with the center of the chamber. We measured the velocity for flow
areas from $60 \times 60$ mm$^2$ up to $212 \times 212$ mm$^2$
with a spatial resolution of $1024 \times 1024$ pixels. This size
of the probed area corresponds to a spatial resolution from 58
$\mu$m~/~pixel up to 207 $\mu$m~/~pixel. These regions were
analyzed with interrogation windows of $32 \times 32$ and $16
\times 16$ pixels. A velocity vector was determined in every
interrogation window, allowing us to construct a velocity map
comprising $32 \times 32$ or $64 \times 64$ vectors. The velocity
maps were determined in two planes. In the one-grid experiments
the plane was parallel to the grid, while in the two-grid
experiments the plane was normal to the grids. The mean and r.m.s.
velocities for each point of the velocity map (1024 points) were
determined by averaging over 100 independent maps, and then over
1024 points. The two-point correlation functions of the velocity
field were determined for each  point of the central part of the
velocity map ($16 \times 16$ vectors) by averaging over 100
independent velocity maps, and then over 256 points. An integral
scale $L$ of turbulence was determined from the two-point
correlation functions of the velocity field. These measurements
were repeated for different distances from the grid, and for
various temperature gradients, Reynolds numbers and particle mass
loadings. The PIV measurements performed in the oscillating grids
turbulence generator confirmed earlier results (Thompson and
Turner 1975; Hopfinger and Toly 1976; Kit et al. 1997) for the
dependence of varies characteristics of the turbulent velocity
field on the parameters of the oscillating grids turbulence
generator.

To determine the spatial distribution of the particles, we used an
Image Processing Technique based on Mie scattering. The advantages
of this method have been demonstrated by Guibert et al. (2001).
The light radiation energy flux scattered by small particles is $
E_s \propto \tilde E \, \Psi(\pi d_p/\lambda; a; n) $, where $
\tilde E \propto \pi d_p^2 / 4$ is the energy flux incident at the
particle, $d_p$ is the particle diameter, $\lambda$ is the
wavelength, $a$ is the index of refraction and $\Psi$ is the
scattering function. In the general case, $\Psi$ is given by Mie
equations. For wavelengths $\lambda$, which are smaller than the
particle size, $\Psi$ tends to be independent of $d_p$ and
$\lambda$. The scattered light energy flux incident on the CCD
camera probe is proportional to the particle number density $n$,
i.e., $ E_s \propto \tilde E \, n (\pi d_p^2 / 4) $.

Mie scattering does not change from temperature effects since it
depends on the permittivity of particles, the particle size and
the laser light wavelength. The effect of the temperature on these
characteristics is negligibly small. Note that in each experiment
before taking the measurements, we let the system run for some
time (up to 30 minutes after smoke injection into the flow with a
steady mean temperature profile) in order to attain a stationary
state.

We found that the probability density function of the particle
size measured with the PM300 particulate analyzer was to the most
extent independent of the location in the flow for incense
particle size of $0.5-1 \, \mu$m. Note that since the number
density of the particles is small (about $1$ mm apart), it can be
assumed that a change in particle number density will not affect
their size distribution. Therefore, the ratio of the scattered
radiation fluxes at two points and at the image measured with the
CCD camera remains equal to the ratio of the particle mean number
densities at these two locations.

\section{Experimental detection of turbulent thermal diffusion}
\label{sec:res}

The turbulent flow parameters in the oscillating grids turbulence
generator are: r.m.s. velocity $\sqrt{\langle {\bf u}^2 \rangle} =
4 - 14$ cm/s depending on the frequency of the grid oscillations,
integral scale of turbulence $L = 1.6 - 2.3$ cm, and the
Kolmogorov length scale $\eta = 380 - 600 \, \mu$m. Other
parameters are given in Section 5 (Tables 1-2). These flows
involve a wide range of scales of turbulent motions.
Interestingly, a flow with a wide range of spatial scales is
already formed at comparatively low frequencies of grid
oscillations. We found a weak mean flow in the form of two large
toroidal structures parallel and adjacent to the grids. The
interaction of these structures results in a symmetric mean flow
that is sensitive to the parameters of the grid adjustments. We
particularly studied the parameters that affect a mean flow such
as the grid distance to the walls of the chamber and partitions,
and the angles between the grid planes and the axes of their
oscillations. Varying these parameters allowed us to expand the
central region with a homogeneous turbulence. We found that the
measured r.m.s. velocity was several times higher than the mean
velocity in the core of the flow.

The temperature measurements were taken in the oscillating grids
turbulence generator. The temperature difference between the heat
exchangers varied within a range of 25 to 50 K. The mean
temperature vertical profiles at a frequency of grids oscillations
$f = 10.5$ Hz in turbulent flows with downward and upward mean
temperature gradients are shown in Fig.~3. Here $Z$ is a
dimensionless vertical coordinate measured in units of the height
of the chamber, and $Z=0$ at the bottom of the chamber. Hereafter,
we use the following system of coordinates: $Z$ is the vertical
axis, the $Y$-axis is perpendicular to the grids and the
$XZ$-plane is parallel to the grids plane.

\begin{figure}
\vspace*{2mm}
\centering
\includegraphics[width=8cm]{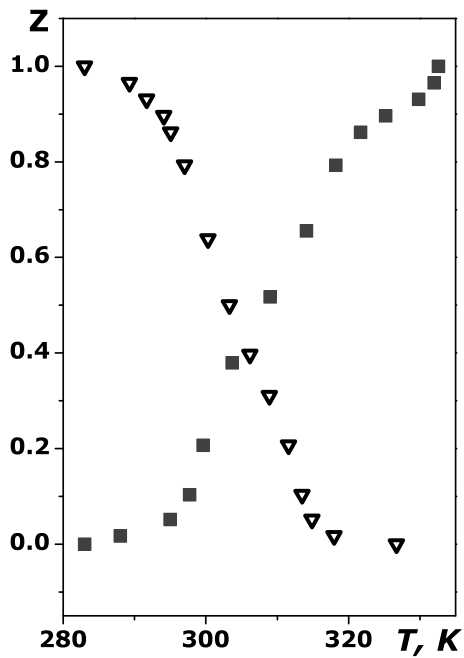}
\caption{\label{Fig3}Vertical temperature profiles at frequency of
grid oscillations $f = 10.5$ Hz in turbulent flows with an upward
mean temperature gradient (filled squares) and with a downward
mean temperature gradient (unfilled triangles). Here $Z$ is a
dimensionless vertical coordinate measured in units of the height
of the chamber, and $Z=0$ at the bottom of the chamber.}
\end{figure}

Measurements performed using different concentrations of incense
smoke showed that  in an isothermal flow the distribution of the
average scattered light intensity over a vertical coordinate is
independent of the mean particle number density. In order to
characterize the spatial distribution of particle number density $
\bar N \propto E^T / E $ in a non-isothermal flow, the
distribution of the scattered light intensity $E$ for the
isothermal case was used to normalize the scattered light
intensity $E^T$ obtained in a non-isothermal flow under the same
conditions. The scattered light intensities $E^T$ and $E$ in each
experiment were normalized by corresponding scattered light
intensities averaged over the vertical coordinate. The ratios $E^T
/ E$ of the normalized average distributions of the intensity of
scattered light as a function of the normalized vertical
coordinate $Z$ in turbulent flows with a downward mean temperature
gradient and with an upward mean temperature gradient are shown in
Fig.~4. The figure demonstrates that particles are redistributed
in a turbulent flow with a mean temperature gradient. Particles
accumulate in regions of minimum mean temperature (in the lower
part of the chamber in turbulent flows with an upward mean
temperature gradient), and there is an outflow of particles from
the upper part of the chamber where the mean temperature is
larger. On the other hand, in a flow with a downward mean
temperature gradient (a hot bottom and a cold top wall of the test
section) particles accumulate in the vicinity of the top wall of
the chamber, i.e., in the vicinity where the mean fluid
temperature is minimum.

\begin{figure}
\centering
\includegraphics[width=8cm]{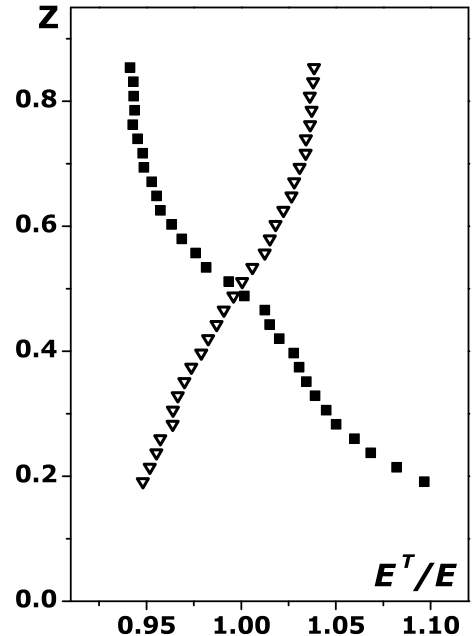}
\caption{\label{Fig4} Ratios $E^T / E$ of normalized average
distributions of the intensity of scattered light versus the
normalized vertical coordinate $Z$ in turbulent flows with an
upward mean temperature gradient (filled squares) and with a
downward mean temperature gradient (unfilled triangles). Here
$Z=0$ is at the bottom of the chamber. The frequency of grids
oscillations is $f = 10.5$ Hz.}
\end{figure}

To determine the turbulent thermal diffusion ratio, we plotted the
normalized particle number density $N_r \equiv \bar N / \bar N_0$
versus the normalized temperature difference $ T_r \equiv (\bar T
- \bar T_0) / \bar T_0 ,$ where $\bar T_0$ is the reference
temperature and $ \bar N_0 = \bar N(\bar T = \bar T_0).$ The
function $N_r(T_r)$ is shown in Fig.~5 for a turbulent flow with
an upward mean temperature gradient and in Fig.~6 for a turbulent
flow with a downward mean temperature gradient.

\begin{figure}
\centering
\includegraphics[width=8cm]{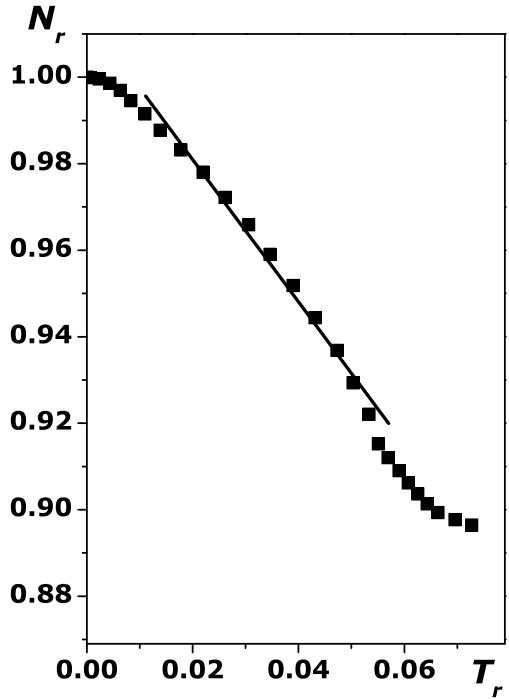}
\caption{\label{Fig5} Normalized particle number density $N_r
\equiv \bar N / \bar N_0$ versus normalized temperature difference
$T_r \equiv (\bar T - \bar T_0) / \bar T_0 $ in a turbulent flow
with an upward mean temperature gradient. The frequency of grid
oscillations is $f = 10.5$ Hz.}
\end{figure}

\begin{figure}
\centering
\includegraphics[width=8cm]{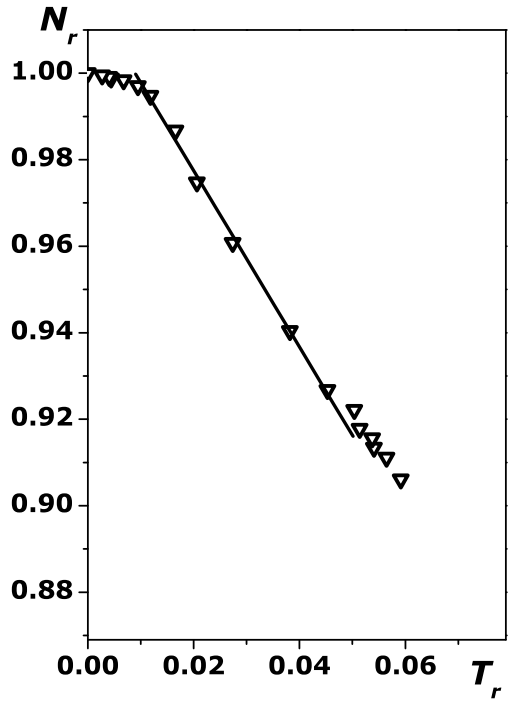}
\caption{\label{Fig6} Normalized particle number density $N_r
\equiv \bar N / \bar N_0$ versus normalized temperature difference
$T_r \equiv (\bar T - \bar T_0) / \bar T_0 $ in a turbulent flow
with a downward mean temperature gradient. The frequency of grid
oscillations is $f = 10.4$ Hz.}
\end{figure}

We probed the central $ 20 \times 20 $ cm region in the chamber by
determining the mean intensity of scattered light in $ 32 \times
16 $ interrogation windows with a size of $ 32 \times 64 $ pixels.
The vertical distribution of the intensity of the scattered light
was determined in 16 vertical strips, which are composed of 32
interrogation windows. Variations of the obtained vertical
distributions between these strips were very small. Therefore in
Figs.~4-6 we used spatial average over strips and the ensemble
average over 100 images of the vertical distributions of the
intensity of scattered light. Figures~5-6 were plotted using the
mean temperature vertical profiles shown in Fig.~3. The normalized
local mean temperatures [the relative temperature differences $
(\bar T - \bar T_0) / \bar T_0 $] in Figs.~5-6 correspond to the
different points inside the probed region. In particular, in
Fig.~5 the location of the point with reference temperature $\bar
T_0$ is $Z=0.135$ (the lowest point of the probed region with a
maximum $\bar N$ in turbulent flows with an upward mean
temperature gradient), and the point with a maximum normalized
temperature difference is the highest point of the probed region
at $Z=0.865 $. The size of the probed region did not affect our
results.

\section{Discussion}
\label{sec:disc}

The vertical profiles of the mean number density of particles
obtained in our experiments can be explained by considering
Eqs.~(\ref{A2}) and~(\ref{A3}). If one does not take into account
the term $\bar N {\bf V}_{\rm eff}$ in Eq.~(\ref{A2}) for the mean
number density of particles, the equation reads $\partial \bar N /
\partial t = D_{_{T}} \Delta \bar N ,$ where we neglected a small
mean fluid velocity $\bar{\bf V}_{p}$ and a small molecular mean
flux of particles (that corresponds to the conditions of the
experiment). The steady-state solution of this equation is $\bar
N= \, $ const. However, our measurements demonstrate that the
solution $\bar N= \, $ const is valid only for an isothermal
turbulent flow, i.e., when $\bar T= \, $ const. By taking into
account the effect of turbulent thermal diffusion, i.e., the term
$\bar N {\bf V}_{\rm eff}$ in Eq.~(\ref{A2}), the steady-state
solution of this equation for non-inertial particles is:
$\bec{\nabla} \bar N / \bar N = - \bec{\nabla} \bar T / \bar T .$
The latter equation yields $\bar N / \bar N_0 = 1 - (\bar T - \bar
T_0) / \bar T_0 ,$ where $ \bar T - \bar T_0 \ll \bar T_0 .$ In
the experiments using particles of the size $0.5 - 2 \, \mu$m, we
found that
\begin{eqnarray}
{\bar N \over \bar N_0} = 1 - \alpha {\bar T - \bar T_0 \over \bar
T_0} \;, \label{R10}
\end{eqnarray}
where the coefficient $\alpha = 1.33 - 1.79 $ (depending on the
frequency of the grid oscillations) in a turbulent flow with an
upward mean temperature gradient, and $\alpha = 1.29 - 1.87 $ in a
turbulent flow with a downward mean temperature gradient (see
Tables 1-2). The deviation of coefficient $\alpha$ from $1$ is
caused by a small yet finite inertia of the particles and by the
dependence of coefficient $\alpha$ on the mean temperature
gradient. The exact value of parameter $\alpha$ for inertial
particles cannot be found within the framework of the theory of
turbulent thermal diffusion (Elperin et al. 1996; 1997; 1998;
2000b; 2001) for the conditions of our experiments (i.e., for
strong mean temperature gradients). However, in all the present
experiments performed for different ranges of parameters and
different directions of a mean temperature gradient, the
coefficient of turbulent thermal diffusion $\alpha$ was more than
$1$, which does agree with the theory. In particular, the theory
predicts $\alpha= 1$ for noninertial particles, and $\alpha > 1$
for inertial particles. As can be seen in Tables 1-2, parameter
$\alpha$ slightly decreases when the Reynolds number increases.

\begin{table}
\begin{tabular}{|c|c|c|c|c|}
\hline
     &      &     &     &       \\
$f \,\,$ (Hz) & $\sqrt{\langle {\bf u}^2 \rangle}\,\,$
(cm/s) & $L \,\,$ (cm) & ${\rm Re}$ & $\alpha$ \\
     &      &     &     &       \\
\hline
     &      &     &     &       \\
6.5  & 3.6  & 1.9 & 46  & 1.79  \\
     &      &     &     &       \\
10.5 & 7.2  & 2.1 & 101 & 1.65  \\
     &      &     &     &       \\
14.5 & 10.7 & 2.3 & 164 & 1.43  \\
     &      &     &     &       \\
16.5 & 12.4 & 2.0 & 165 & 1.33  \\
     &      &     &     &       \\
\hline
\end{tabular}
\caption[]{Parameters of turbulence and the turbulent thermal
diffusion coefficient for a turbulent flow with an upward mean
temperature gradient. Here ${\rm Re}=L \, \sqrt{\langle {\bf u}^2
\rangle} / \nu$, $L$ is the integral scale of turbulence, and
$\nu$ is the kinematic viscosity.}
\end{table}

\begin{table}
\begin{tabular}{|c|c|c|c|c|}
\hline
     &      &     &     &       \\
$f \,\,$ (Hz) & $\sqrt{\langle {\bf u}^2 \rangle}\,\,$
(cm/s) & $L \,\,$ (cm) & ${\rm Re}$ & $\alpha$ \\
     &      &     &     &       \\
\hline
     &      &     &        &       \\
8.4  & 8.8  & 1.85 & 109   & 1.87  \\
     &      &      &       &       \\
10.4 & 9.7  & 1.75 & 113   & 1.60  \\
     &      &      &       &       \\
12.4 & 11.1 & 1.65 & 122   & 1.69  \\
     &      &      &       &       \\
14.4 & 11.7 & 1.84 & 144   & 1.34  \\
     &      &      &       &       \\
16.4 & 14.0 & 1.64 & 153   & 1.29  \\
     &      &      &       &       \\
\hline
\end{tabular}
\caption[]{Parameters of turbulence and the turbulent thermal
diffusion coefficient for a turbulent flow with a downward mean
temperature gradient.}
\end{table}

There are other factors that can affect the spatial distribution
of particles. The contribution of the mean flow to the spatial
distribution of particles is negligibly small. Indeed, the
normalized distribution of the scattered light intensity measured
in the different vertical strips in the regions where the mean
flow velocity and the coefficient of turbulent diffusion vary
strongly are practically identical (the difference being only
about 1 \%). Due to the effect of turbulent thermal diffusion,
particles are redistributed in the vertical direction in the
chamber. In turbulent flows with an upward mean temperature
gradient particles accumulated in the lower part of the chamber,
and in flows with a downward mean temperature gradient particles
accumulated in the vicinity of the top wall of the chamber, i.e.,
in regions with a minimum mean temperature. The spatial-temporal
evolution of the normalized number density of particles $N_r$ is
governed by the conservation law of the total number of particles
(see Eqs.~(\ref{A2}) and~(\ref{A3})). Some fraction of particles
sticks to the walls of the chamber, and the total number of
particles without feeding fresh smoke slowly decreases. The
characteristic time of this decrease is about 15 minutes. However,
the spatial distribution of the normalized number density of
particles does not change over time.

The number density of particles in our experiments was of the
order of $10^{10}$ particles per cubic meter. Therefore, the
distance between particles is about $1$ mm, and their collision
rate is negligibly small. Indeed, calculation of the particle
collision rate using the Saffman-Turner formula (Saffman and
Turner 1956) confirmed this finding.  Consequently, we did not
observe any coalescence of particles. The effect of the
gravitational settling of small particles ($0.5 - 1 \, \mu$m) is
negligibly small (the terminal fall velocity of these particles
being less than $0.01$ cm/s). Thus, we may conclude that the two
competitive mechanisms of particle transport, i.e., mixing by
turbulent diffusion and accumulation of particles due to turbulent
thermal diffusion, exist simultaneously and there is a very small
effect of gravitational settling of the particles.

In regard to the accuracy of the performed measurements, we found
that uncertainties in optics adjustment and errors in measuring a
CCD image background value are considerably less than the observed
effect of a $10 \%$ change of normalized intensity of the
scattered light in the test section in the presence of an imposed
mean temperature gradient. In the range of the tracer
concentrations used in the experiments the particle-air suspension
can be considered as an optically thin medium, from which we can
infer that the intensity of the scattered light is proportional to
the particle number density. The total error in our measurements
was determined by $(\delta \bar N / \bar N + \delta \bar T / \bar
T) / \sqrt{Q} \approx 0.3 \% $ (see Eq.~(\ref{R10})), where
$\delta \bar T = 0.1 K$ is the accuracy of the temperature
measurements; $\delta \bar N / \bar N = 0.8 \%$ is the accuracy of
the mean number density measurements; and $Q=8$ is the number of
experiments performed for each direction of the mean temperature
gradient and for each value of the frequency of the grid
oscillations. The total variation of the normalized particle
number density due to the effect of turbulent thermal diffusion is
more than $10 \%$ (see Figs.~4-6). The relative error of the
measurements is less than $3 \%$ of the total variation of the
particle number density. Thus, the accuracy of these measurements
is considerably higher than the magnitude of the observed effect.
As such, our experiments confirm the existence of an effect of
turbulent thermal diffusion, as predicted theoretically by Elperin
et al. (1996; 1997).

\section{Conclusions}
\label{sec:end}

A new phenomenon of turbulent thermal diffusion has been detected
experimentally in turbulence generated by oscillating grids with
an imposed vertical mean temperature gradient in air flow. This
phenomenon implies that there exists an additional mean flux of
particles in the direction opposite to that of the mean
temperature gradient, that results in the formation of large-scale
inhomogeneities in the spatial distribution of particles. The
particles accumulated in the vicinity of the minimum mean fluid
temperature. In the experiments in two directions of the vertical
mean temperature gradient, it was found that in a flow with a
downward mean temperature gradient particles accumulate in the
vicinity of the top wall of the chamber. In a flow with an upward
mean temperature gradient particles accumulate in the vicinity of
the bottom wall of the chamber. Turbulent thermal diffusion can
explain the large-scale aerosol layers that form inside
atmospheric temperature inversions.

\begin{acknowledgements}
We are indebted to F.~Busse, H.~J.~S.~Fernando, J.~Katz, E.~Kit,
V.~L'vov, J.~Mann, S.~Ott and A.~Tsinober for illuminating
discussions. We also thank A. Markovich for his assistance in
processing the experimental data. This work was partially
supported by the German-Israeli Project Cooperation (DIP)
administrated by the Federal Ministry for Education and Research
(BMBF) and by the Israel Science Foundation governed by the
Israeli Academy of Science.
\end{acknowledgements}

\end{document}